\documentclass[12pt]{article}
\usepackage[utf8]{inputenc}
\usepackage{geometry}
\usepackage{amsmath}

\usepackage[table,dvipsnames]{xcolor}

\usepackage{tikz}
\usepackage{graphicx}
\usepackage[hidelinks]{hyperref}
\usepackage{natbib}

\title{When Numbers Mislead Us}
\author{Arthur Charpentier}
\date{Université du Québec à Montréal \\ charpentier.arthur@uqam.ca}

\begin{document}

\maketitle

Believing that there is a single, objective way to describe phenomena using numbers is to forget that data does not “speak” for itself. Collecting data involves making choices: what to measure, how, when, on whom, etc. This implies implicit (or even ideological) assumptions about what counts as a measurable fact. And in any data analysis, what is not measured can be as important as what is observed. When an influential variable is overlooked—whether ignored, neglected, or simply unknown—the apparent relationships between other variables can become misleading. This is known as “omitted variable bias”: a hidden effect distorts comparisons and can make a correlation appear where there is none, or mask a real one. Sometimes, introducing this “forgotten” variable can even completely reverse the conclusions that would have been drawn from a naive reading of the data.

\section{Rankings Hospitals, Doctors, Schools, etc.}

In an era of widespread rankings—from the best hospitals to the best doctors, including schools, as frequently found in the press—performance indicators are playing an increasingly important role in public decisions and social perceptions. But as Goodhart's law states, “when a measure becomes a target, it ceases to be a good measure.” When a hospital seeks to climb the rankings, it may do so not by actually improving the quality of care, but by acting on the levers that influence the indicators: admitting less serious patients, avoiding complex cases, or optimizing its discharge statistics. The risk is then to produce a biased image of efficiency, disconnected from medical reality. Consider the following (entirely fictional) example, from Table \ref{tab:hospital}, with two hospitals of the same size, very loosely based on a famous example of kidney stone treatment (with real data) that appeared in the British Medical Journal in the 1980s. Based, on a quick comparison of mortality rates in two hospitals, Hospital A has better outcomes within both subgroups (healthy and non-healthy patients), its overall death rate is higher than Hospital B's.

\begin{table}[!ht]

\begin{minipage}{.5\textwidth}
   \begin{center}

{\renewcommand{\arraystretch}{1.25}
\begin{tabular}{lccccccc}
            & \multicolumn{3}{c}{\cellcolor{BrickRed!10}{\textcolor{BrickRed}{hospital A}}}                                                                               & & \multicolumn{3}{c}{\cellcolor{NavyBlue!10}{\textcolor{NavyBlue}{hospital B}}}                             \\                                                          & \cellcolor{BrickRed!10}{\color{BrickRed} total} & \cellcolor{BrickRed!10}{\color{BrickRed} death} & \cellcolor{BrickRed!10}{\color{BrickRed} ratio} & &\cellcolor{NavyBlue!10}{\color{NavyBlue} total} & \cellcolor{NavyBlue!10}{\color{NavyBlue} death} & \cellcolor{NavyBlue!10}{\color{NavyBlue} ratio} \\\arrayrulecolor{white}\hline
            non-healthy & \cellcolor{BrickRed!10}{\color{BrickRed} {60}}     & \cellcolor{BrickRed!10}{\color{BrickRed} {36}}     & \cellcolor{BrickRed!10}{\color{BrickRed} 60\%}   & $<$ &\cellcolor{NavyBlue!10}{\color{NavyBlue} 20}     & \cellcolor{NavyBlue!10}{\color{NavyBlue} 14}     & \cellcolor{NavyBlue!10}{\color{NavyBlue} 70\%}      \\
            healthy & \cellcolor{BrickRed!10}{\color{BrickRed} 20}     & \cellcolor{BrickRed!10}{\color{BrickRed} 4}     & \cellcolor{BrickRed!10}{\color{BrickRed} 20\%}   & $<$ &\cellcolor{NavyBlue!10}{\color{NavyBlue} 60}      & \cellcolor{NavyBlue!10}{\color{NavyBlue} 18}      & \cellcolor{NavyBlue!10}{\color{NavyBlue} 30\%}      \\\arrayrulecolor{white}\hline
     total & \cellcolor{BrickRed!10}{\color{BrickRed} 80}   & \cellcolor{BrickRed!10}{\color{BrickRed} 40}   & \cellcolor{BrickRed!10}{\color{BrickRed} 50\%}      & $>$ &\cellcolor{NavyBlue!10}{\color{NavyBlue} 80}      & \cellcolor{NavyBlue!10}{\color{NavyBlue} 32}      & \cellcolor{NavyBlue!10}{\color{NavyBlue} 40\%}      \\

\end{tabular}}
\end{center}
\end{minipage}
\caption{Mortality rates by hospital and patient condition (source: author).}
    \label{tab:hospital}
\end{table}

\section{Not Really a Paradox...}

In 1951, British statistician Edward Simpson published an article on contingency table analysis, in \cite{simpson1951interaction}. In it, he described a phenomenon whereby the relationship between two variables changes when a third hidden variable is taken into account. He did not consider this a paradox, but rather a statistical effect worth noting. It was not until the 1970s that this behavior was rediscovered, named in his honor, and elevated to the status of a “paradox” because it defies our intuition.
Despite its name, Simpson's paradox is not a paradox in the mathematical sense, as it is not based on any logical contradiction or error in formal reasoning. Rather, it is a statistical surprise linked to our intuition about averages. In reality, this phenomenon occurs when ratios calculated in subgroups (e.g., recovery rates in two categories of patients) are compared and then aggregated without taking into account the structure of the data (group size, distribution of cases). The misinterpretation stems from the fact that the overall ratio is not the average of the ratios: in general, percentages from groups of different sizes cannot be added together and used to derive a meaningful average without weighting them. In the previous example, for Hospital A, the total death rate is not the average of the two rates (otherwise we would have obtained 40\%, the average of 60\% and 20\%) but 50\%. And then we have a small counterintuitive property about fractions, namely that we can have
$$
\textcolor{BrickRed}{\frac{36}{60}}<
\textcolor{NavyBlue}{\frac{14}{20}}\text{ and }
\textcolor{BrickRed}{\frac{4}{20}}<
\textcolor{NavyBlue}{\frac{18}{60}}
$$
but, at the same time,
$$
\textcolor{BrickRed}{\frac{36+4}{60+20}}=\textcolor{BrickRed}{\frac{40}{80}}>
\textcolor{NavyBlue}{\frac{14+18}{20+60}}=
\textcolor{NavyBlue}{\frac{32}{80}}
$$

More formally, in Table \ref{tab:hospital:abcd}, we have an abstract representation of the paradox (inspired from Table \ref{tab:hospital}). Each group (A and B) is split into two subgroups. Even though group B shows worse ratios in both subgroups, its overall rate is better—due to differences in the distribution of totals across groups.

A graphical depiction of the paradox: line slopes represent subgroup ratios. In both cases, group B's ratios are higher, but once aggregated, group A’s overall slope becomes steeper—demonstrating the paradox in action.

\begin{figure}[]
    \centering

    \begin{minipage}{.5\textwidth}
\begin{center}
\begin{tikzpicture}[scale=1.4]

\coordinate (O) at (0,0);
\coordinate (B1) at (0.2223, 0.5958);
\coordinate (R1) at (0.8443, 1.4498);
\coordinate (R2) at (1.8687, 0.4402);
\coordinate (B2) at (2.5047, 0.9260);
\coordinate (B3) at (2.6848, 1.5142);
\coordinate (R3) at (2.7214, 1.9014);
\coordinate (B1b) at (0.2223, 0.68);
\coordinate (R1b) at (0.8443, 1.4498);
\coordinate (R2b) at (1.8687, 0.4402);
\coordinate (B2b) at (2.5047, 0.9260);
\coordinate (B3b) at (2.6848, 1.5142);
\coordinate (R3b) at (2.7214, 1.9014);

% Axes
\draw[->, thick] (0,0) -- (3.8,0) node[right] {$x$ (total)};
\draw[->, thick] (0,0) -- (0,2) node[above] {$y$ (deaths)};

% Red line
\draw[BrickRed, thick, dashed] (O) -- (R1);
\draw[BrickRed, thick, dashed] (O) -- (R2);
\draw[BrickRed, thick] (O) -- (R3);
\draw[NavyBlue, thick, dashed] (O) -- (B1);
\draw[NavyBlue, thick, dashed] (O) -- (B2);
\draw[NavyBlue, thick] (O) -- (B3);
\draw[BrickRed, thick, dashed] (R3) -- (R1);
\draw[BrickRed, thick, dashed] (R3) -- (R2);
\draw[NavyBlue, thick, dashed] (B3) -- (B1);
\draw[NavyBlue, thick, dashed] (B3) -- (B2);

\filldraw[BrickRed] (R1) circle (2pt) node[above] {$(a,b)$};
\filldraw[BrickRed] (R2) circle (2pt) node[right] {$(c,d)$};
\filldraw[BrickRed] (R3) circle (2pt) node[above right] {$(a+c,b+d)$};
\filldraw[NavyBlue] (B1) circle (2pt) node[above] {$(A,B)$};
\filldraw[NavyBlue] (B2) circle (2pt) node[right] {$(C,D)$};
\filldraw[NavyBlue] (B3) circle (2pt) node[right] {$(A+C,B+D)$};

% Origine
\filldraw[black] (0,0) circle (2.5pt);
\end{tikzpicture}
\end{center}
\end{minipage}
\caption{Geometric visualization of Simpson’s paradox, based on Figure \ref{tab:hospital:abcd}.}
    \label{fig:hospital}
\end{figure}
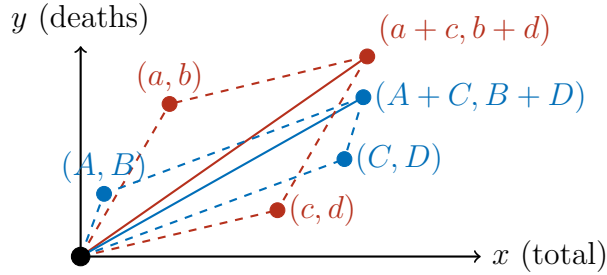

\begin{table}
\begin{minipage}{.5\textwidth}
   \begin{center}

{\renewcommand{\arraystretch}{1.25}
\hspace{-2cm}\begin{tabular}{lccccccc}
            & \multicolumn{3}{c}{\cellcolor{BrickRed!10}{\textcolor{BrickRed}{group A}}}                                                                               & & \multicolumn{3}{c}{\cellcolor{NavyBlue!10}{\textcolor{NavyBlue}{group B}}}                             \\                                                          & \cellcolor{BrickRed!10}{\color{BrickRed} total} & \cellcolor{BrickRed!10}{\color{BrickRed} deaths} & \cellcolor{BrickRed!10}{\color{BrickRed} ratio} & &\cellcolor{NavyBlue!10}{\color{NavyBlue} total} & \cellcolor{NavyBlue!10}{\color{NavyBlue} deaths} & \cellcolor{NavyBlue!10}{\color{NavyBlue} ratio} \\\arrayrulecolor{white}\hline
            subgroup 0 & \cellcolor{BrickRed!10}{\color{BrickRed} $a$}     & \cellcolor{BrickRed!10}{\color{BrickRed}$b$}     & \cellcolor{BrickRed!10}{\color{BrickRed} $\frac{\displaystyle b}{\displaystyle a}$}   & $<$ &\cellcolor{NavyBlue!10}{\color{NavyBlue} $A$}     & \cellcolor{NavyBlue!10}{\color{NavyBlue} $B$}     & \cellcolor{NavyBlue!10}{\color{NavyBlue} $\frac{\displaystyle B}{\displaystyle A}$}      \\
            subgroup 1 & \cellcolor{BrickRed!10}{\color{BrickRed} $c$}     & \cellcolor{BrickRed!10}{\color{BrickRed} $d$}     & \cellcolor{BrickRed!10}{\color{BrickRed} $\frac{\displaystyle d}{\displaystyle c}$}   & $<$ &\cellcolor{NavyBlue!10}{\color{NavyBlue} $C$}      & \cellcolor{NavyBlue!10}{\color{NavyBlue} $D$}      & \cellcolor{NavyBlue!10}{\color{NavyBlue} $\frac{\displaystyle D}{\displaystyle C}$}      \\\arrayrulecolor{white}\hline
     total & \cellcolor{BrickRed!10}{\color{BrickRed} $a+c$}   & \cellcolor{BrickRed!10}{\color{BrickRed} $b+d$}   & \cellcolor{BrickRed!10}{\color{BrickRed} $\frac{\displaystyle b+d}{\displaystyle a+c}$}      & $>$ &\cellcolor{NavyBlue!10}{\color{NavyBlue} $A+C$}      & \cellcolor{NavyBlue!10}{\color{NavyBlue} $B+D$}      & \cellcolor{NavyBlue!10}{\color{NavyBlue} $\frac{\displaystyle B+D}{\displaystyle A+C}$}      \\

\end{tabular}}
\end{center}
\end{minipage}
\caption{Formal version of Table \ref{tab:hospital}, with counts, by hospital and patient condition.}\label{tab:hospital:abcd}
\end{table}

xxx
$$
\textcolor{BrickRed}{\frac{b}{a}}<
\textcolor{NavyBlue}{\frac{B}{A}}\text{ and }
\textcolor{BrickRed}{\frac{d}{c}}<
\textcolor{NavyBlue}{\frac{D}{C}}
$$
but, at the same time,
$$
\textcolor{BrickRed}{\frac{b+d}{a+c}}>
\textcolor{NavyBlue}{\frac{B+D}{A+C}}
$$
These inequalities can be seen in Figure \ref{fig:hospital}. The segment connecting (20,4) from the origin (0,0), in red, has a slope of 4/20, which is below the segment connecting (60,18), because its slope of 18/60 is greater. The same applies to the red segment connecting (60,36), which is below the blue segment connecting (20,14) because $14/20>36/60$. However, when we aggregate, we obtain the points on the right, (80,32) in blue and (80,40) in red, and the blue slope is lower than the red slope.

This discrepancy between the overall ratio and the partial ratios fuels many well-known “paradoxes” in mathematics concerning averages: harmonic mean vs. arithmetic mean, average speed, or even the “average of averages.” Simpson's paradox is a sophisticated version of the latter case: it reminds us that a misinterpreted average can not only be misleading, but can completely reverse a trend. It is therefore not a mathematical oddity, but an invitation to caution in the interpretation of aggregates.

\section{Berkeley Admissions and Gender Bias: A Real Case}

A famous case highlighted this pitfall: in 1973, the University of California, Berkeley was accused of discriminating against women in graduate admissions. The overall figures were unequivocal: 44\% of men were admitted, compared to only 35\% of women—an alarming difference that was enough to trigger a lawsuit. This significant difference raised suspicions of gender discrimination in the selection process. However, upon closer examination of the data, something surprising emerged: in most departments, the admission rate for women was equal to or higher than that for men. The data provided by Peter Bickel, Eugene Hammel, Eugene, and William O'Connell shows the same kind of phenomenon observed in hospitals, in \cite{bickel1975sex} or \cite{bickel1977sex}.

\begin{table}[]
\begin{center}
{\renewcommand{\arraystretch}{1.25}
\begin{tabular}{cccccccc}
    & \multicolumn{3}{c}{\cellcolor{BrickRed!10}{\textcolor{BrickRed}{Men}}} & & \multicolumn{3}{c}{\cellcolor{NavyBlue!10}{\textcolor{NavyBlue}{Women}}} \\
    & \cellcolor{BrickRed!10}{\color{BrickRed} Total} & \cellcolor{BrickRed!10}{\color{BrickRed} Admitted} & \cellcolor{BrickRed!10}{\color{BrickRed} Rate}
    & & \cellcolor{NavyBlue!10}{\color{NavyBlue} Total} & \cellcolor{NavyBlue!10}{\color{NavyBlue} Admitted} & \cellcolor{NavyBlue!10}{\color{NavyBlue} Rate} \\\arrayrulecolor{white}\hline
    A & \cellcolor{BrickRed!10}{\color{BrickRed} 825} & \cellcolor{BrickRed!10}{\color{BrickRed} 512} & \cellcolor{BrickRed!10}{\color{BrickRed} 62.1\%}
      & $<$ & \cellcolor{NavyBlue!10}{\color{NavyBlue} 108} & \cellcolor{NavyBlue!10}{\color{NavyBlue} 89} & \cellcolor{NavyBlue!10}{\color{NavyBlue} 82.4\%} \\
    B & \cellcolor{BrickRed!10}{\color{BrickRed} 560} & \cellcolor{BrickRed!10}{\color{BrickRed} 353} & \cellcolor{BrickRed!10}{\color{BrickRed} 60.0\%}
      & $<$ & \cellcolor{NavyBlue!10}{\color{NavyBlue} 25} & \cellcolor{NavyBlue!10}{\color{NavyBlue} 17} & \cellcolor{NavyBlue!10}{\color{NavyBlue} 68.0\%} \\
    C & \cellcolor{BrickRed!10}{\color{BrickRed} 325} & \cellcolor{BrickRed!10}{\color{BrickRed} 120} & \cellcolor{BrickRed!10}{\color{BrickRed} 36.9\%}
      & $>$ & \cellcolor{NavyBlue!10}{\color{NavyBlue} 593} & \cellcolor{NavyBlue!10}{\color{NavyBlue} 202} & \cellcolor{NavyBlue!10}{\color{NavyBlue} 34.0\%} \\
    D & \cellcolor{BrickRed!10}{\color{BrickRed} 417} & \cellcolor{BrickRed!10}{\color{BrickRed} 138} & \cellcolor{BrickRed!10}{\color{BrickRed} 33.1\%}
      & $<$ & \cellcolor{NavyBlue!10}{\color{NavyBlue} 375} & \cellcolor{NavyBlue!10}{\color{NavyBlue} 131} & \cellcolor{NavyBlue!10}{\color{NavyBlue} 34.9\%} \\
    E & \cellcolor{BrickRed!10}{\color{BrickRed} 191} & \cellcolor{BrickRed!10}{\color{BrickRed} 53} & \cellcolor{BrickRed!10}{\color{BrickRed} 27.7\%}
      & $>$ & \cellcolor{NavyBlue!10}{\color{NavyBlue} 393} & \cellcolor{NavyBlue!10}{\color{NavyBlue} 94} & \cellcolor{NavyBlue!10}{\color{NavyBlue} 23.9\%} \\
    F & \cellcolor{BrickRed!10}{\color{BrickRed} 373} & \cellcolor{BrickRed!10}{\color{BrickRed} 22} & \cellcolor{BrickRed!10}{\color{BrickRed} 5.9\%}
      & $<$ & \cellcolor{NavyBlue!10}{\color{NavyBlue} 341} & \cellcolor{NavyBlue!10}{\color{NavyBlue} 24} & \cellcolor{NavyBlue!10}{\color{NavyBlue} 7.0\%} \\\arrayrulecolor{white}\hline
    Total & \cellcolor{BrickRed!10}{\color{BrickRed} 2691} & \cellcolor{BrickRed!10}{\color{BrickRed} 1198} & \cellcolor{BrickRed!10}{\color{BrickRed} 44.5\%}
      & $>$ & \cellcolor{NavyBlue!10}{\color{NavyBlue} 1835} & \cellcolor{NavyBlue!10}{\color{NavyBlue} 557} & \cellcolor{NavyBlue!10}{\color{NavyBlue} 30.4\%} \\
\end{tabular}
}
\end{center}

\caption{UC Berkeley graduate admissions, from \cite{bickel1975sex} (real admission data).}
    \label{tab:berkeley}
\end{table}

Table \ref{tab:berkeley} reveals that in four of the six departments, women had a higher admission rate than men. Nevertheless, their overall admission rate was lower—a seemingly contradictory outcome. Why? Because women applied more to the most competitive departments, where admission rates are low for everyone. Men, on the other hand, applied in greater numbers to departments with high admission rates. As a result, the overall rates give an impression that is the opposite of what we get when analyzing each subgroup. Once the data are properly stratified, Bickel and his colleagues even conclude that there is a slight bias in favor of women: “If the data are properly pooled... there is a small but statistically significant bias in favor of women.”
What this example shows is that overall averages can mask structural effects. One might think that women are disadvantaged, when in fact they have simply chosen more competitive fields. This is the crux of Simpson's paradox: it teaches us to be wary of unadjusted overall comparisons. To fully understand a statistical situation, it is often necessary to analyze the data according to relevant subgroups, rather than simply looking at the aggregate figures. This paradox is therefore an excellent lesson in scientific rigor: understanding is not just about observing numbers, it is about knowing what they tell us—and what they hide.

\section{Two More Examples of the Paradox}

\begin{figure}
    \centering
\includegraphics[width=0.95\linewidth]{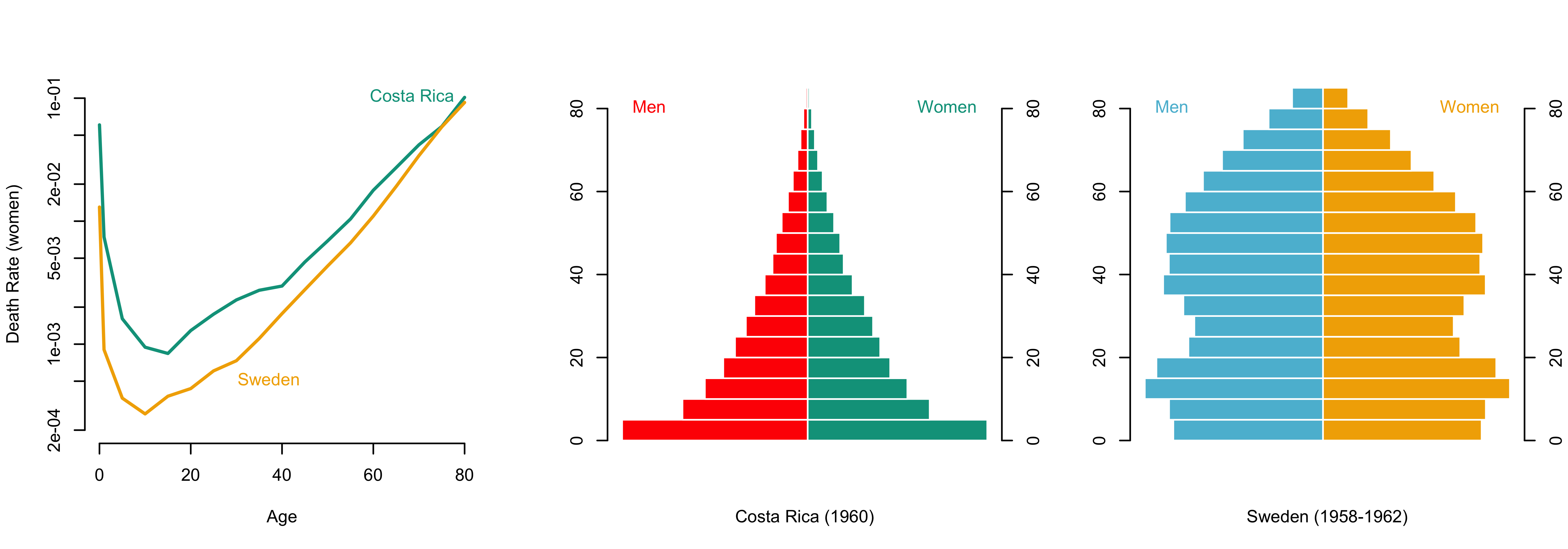}
    \caption{Age-specific mortality rates and population structures in Costa Rica and Sweden, from \cite{charpentier2023springer}. The left panel shows the annual probability of death (on a logarithmic scale) for women by age: at every age, mortality is lower in Sweden than in Costa Rica. The center and right panels display population pyramids from the same period, revealing that Costa Rica had a much younger population, while Sweden had a higher proportion of older individuals. This explains why Sweden's overall mortality rate was higher despite having lower age-specific death rates—a striking example of Simpson’s paradox in demography.}
    \label{fig:suede}
\end{figure}

In 1986, demographer Joel Cohen studied mortality rates in Costa Rica and Sweden—two very different countries, with Sweden renowned for its high life expectancy, in \cite{cohen1986uncertainty}. However, he noted a surprising fact: the overall mortality rate is higher in Sweden (9.29‰) than in Costa Rica (8.12‰). This would be surprising in itself, but another detail reinforces the paradox: in each age group, Sweden has a lower mortality rate than Costa Rica, as shown in Figure \ref{fig:suede}. How is this possible?

The answer lies in the age structure of the populations. Comparing the age pyramids, we find that Costa Rica had a very young population at the time—half of the inhabitants were under 20—while Sweden had a much higher proportion of older people. However, as the probability of dying increases with age, an older population will automatically have a higher overall mortality rate, even if it is healthier at each age. So, if you happen to meet a woman on the street in Costa Rica, she is statistically less likely to be elderly—and therefore less likely to die within the year—than if you meet a woman in Sweden. It is not that Swedish women die more often: it is that the Swedish women you meet are, on average, older. That is the subtlety of the paradox.

In 2004, Gary Davis, a researcher at the University of Minnesota, looked at the relationship between average vehicle speed and the number of accidents involving pedestrians in different neighborhoods of a city, in \cite{davis2004aggregation}. At first glance, the results of his model were puzzling: they seemed to show that reducing the speed limit from 30 to 25 miles per hour led to an increase in the number of accidents. Such a conclusion goes against intuition, the results of controlled experiments, and road safety logic. But upon closer inspection, Davis identified an interpretation error due to data aggregation—a typical effect of Simpson's paradox.
The main problem was that residential areas, which naturally have less traffic and therefore fewer accidents, were also the most likely to adopt the new 25 mph speed limit. As a result, 30 mph streets remained associated with higher traffic volumes. When comparing all streets globally, without taking these contextual differences into account, it appears that the reduction in speed is linked to an increase in accidents — when in reality, it is simply that very different areas are being compared. The model did not take into account this confounding variable: the nature of the neighborhood (residential or non-residential), which determines both traffic volume and accident risk.

\section*{The Ecological Inference Paradox}

Simpson's paradox can be formulated as follows: “{\em what is true in each group may be false in the whole}.” A reverse interpretation is also possible: “{\em what is true at the group level is not necessarily true at the individual level}.” This is often referred to as the “{\em ecological inference paradox},” which originated in an article published in 1950 by American sociologist William S. Robinson (who was the first to use the term “{\em ecological inference paradox},” in \cite{robinson1950ecological}). Robinson was studying the correlation between immigration rates and literacy rates in the United States, comparing data by state. He observed that at the state level, the higher the rate of immigration, the higher the average literacy rate. But when looking at individual data, the opposite was true: immigrants were, on average, less literate than non-immigrants. In other words, the correlations observed at the group level (countries, regions, social classes, etc.) can be very different—even opposite—from those observed at the individual level. The paradox arises when we infer (or draw conclusions) about individuals from aggregated data, which is often a mistake. In this context, the term “ecological” does not refer to environmental ecology, but to analysis by “collective units,” such as social groups, geographical areas, or institutions—as opposed to individual units.

\begin{figure}
    \centering
    \includegraphics[width=0.49\linewidth]{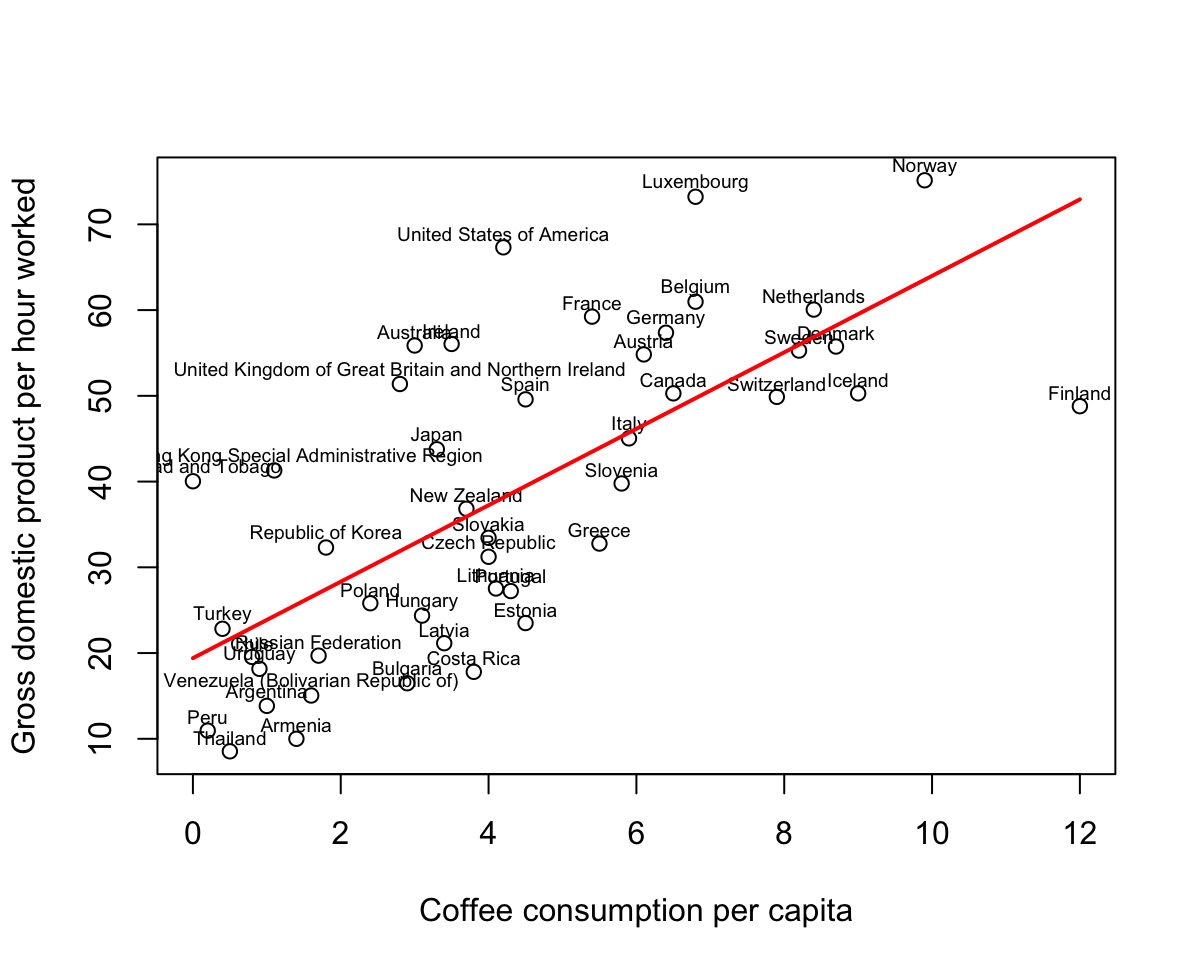} \includegraphics[width=0.49\linewidth]{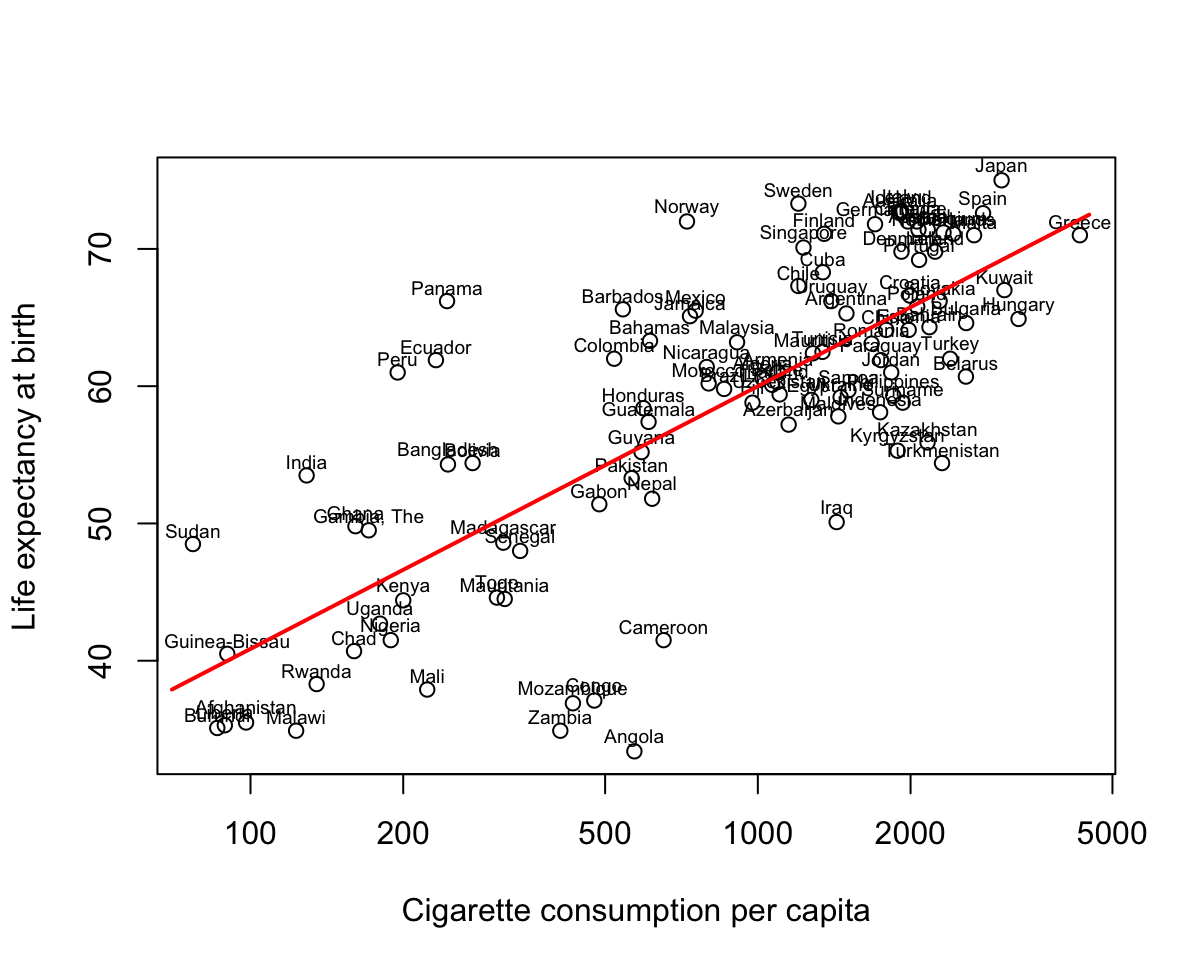}
    \caption{Cigarette consumption vs. life expectancy (left) and coffee consumption vs. labor productivity (right, per country,)}
    \label{fig:coffee:cigarette}
\end{figure}

To illustrate, we can look at the two examples of Figure \ref{fig:coffee:cigarette}, which show, for a large number of countries around the world, gross domestic product per hour worked as a function of per capita coffee consumption (left hand side); or life expectancy at birth as a function of per capita cigarette consumption per year (right hand side).

In both figures, there is a positive correlation at the aggregate level, by country. But it would be risky to claim that the link observed between the variables at the country level reflects a similar relationship at the individual level.
In the first case, the more coffee a country consumes per capita, the higher its GDP per hour worked. Does this suggest that drinking coffee makes you productive? No, it does not mean that an employee who drinks more coffee will be more efficient individually. It is even possible that within a country, heavy coffee drinkers have a comparable or even lower level of productivity than those who do not drink coffee. The explanation is quite simple: coffee is consumed more in rich industrialized countries, where working conditions, automation, technology, and work organization explain productivity. Coffee, here, is just a cultural marker (present in Northern countries, for example), correlated with wealth, but not causal.
In the second case, the more cigarettes a country consumes (per capita), the higher the life expectancy seems to be. This seems to contradict everything we know about the effects of tobacco on health! Individually, smoking reduces life expectancy. Again, in these data, the countries that consume the most cigarettes were also, in general, the richest countries (Japan, France, Greece, etc.), which have better healthcare systems, better nutrition, better living conditions, etc. Conversely, countries with low cigarette consumption are often poorer, and life expectancy there is reduced for structural reasons (access to healthcare, infant mortality, wars, etc.). We therefore observe a positive correlation at the country level, but the individual effect of tobacco remains negative.

\section{Consequences}

These examples remind us that aggregating data can mask or reverse trends. When an important variable is ignored, the overall results become misleading. This is known as omitted variable bias, or confounding variable bias. The consequences are not trivial: they can lead to wrong decisions, or even miscarriages of justice, as Cathleen O'Grady points out in 2023, in \cite{ogrady2023unlucky}, referring to the case of Lucia de Berk (convicted of murdering five children) and the report published by the Royal Statistical Society in 2022 (in \cite{green2022healthcare}). Indeed, some nurses or interns, who mainly care for more seriously ill patients, are statistically associated with more deaths, not because they are at fault, but because their department treats more critical cases. This effect is a classic case of omitted variable bias: the severity of the patients' condition (the hidden variable) influences both the death and the assignment to the caregiver, completely distorting the perception of the mortality rate.

\bibliographystyle{plainnat}
\bibliography{biblio}

\end{document}